\documentclass[a4paper,12pt]{article}

\begin{document}
\makeatletter
\renewcommand{\theequation}{\thesection.\arabic{equation}}
\@addtoreset{equation}{section}
\makeatother

\title{\Large  Kinematical Reduction of Spatial Degrees of Freedom and Holographic Relation in Yang's 
Quantized Space-Time Algebra}
\author{\large Sho Tanaka\footnote{Em. Professor of Kyoto University, E-mail: st-desc@kyoto.zaq.ne.jp }
\\[8 pt]
 Kurodani 33-4, Sakyo-ku, Kyoto 606-8331, Japan}

\date{}
\maketitle

\vspace{-10cm}
\rightline{}
\vspace{12cm}
\thispagestyle{empty}

 We try to find a possible origin of the holographic principle in the Lorentz-covariant Yang's quantized 
space-time algebra (YSTA). YSTA, which is intrinsically equipped with short- and long-scale 
parameters, $\lambda$ and $R$, gives a finite number of spatial degrees of freedom for any bounded 
spatial region, providing a basis for divergence-free quantum field theory. Furthermore, it gives a 
definite kinematical reduction of spatial degrees of freedom, compared with the ordinary lattice 
space.  On account of the latter fact, we find a certain kind of kinematical holographic 
relation in YSTA, which may be regarded as a primordial form of the holographic principle suggested 
so far in the framework of the present quantum theory that appears now in the contraction limit of YSTA, 
$\lambda \rightarrow 0$ and $R \rightarrow \infty.$ 

\vspace{2\baselineskip}

\newpage

\section{\normalsize Introduction}

According to the notice given in the previous paper,$^{[1]}$ we try to find a possible origin of the holographic 
principle in the Lorentz-covariant Yang's quantized space-time algebra (YSTA).$^{[1],[2]}$ 
As was extensively studied in the preceding works,$^{[1],[3],[4]}$  YSTA, which is 
intrinsically equipped with short- and long-scale parameters, $\lambda$ and $R$, gives a finite number 
of spatial degrees of freedom for any finite spatial region and provides a basis for the field theory free 
from ultraviolet- and infrared-divergences. 

Furthermore, we find out below an important fact that YSTA gives a definite kinematical reduction 
of spatial degrees of freedom, compared with the ordinary lattice space. On account of this fact, 
we really find out a kind of kinematical holographic relation. It may be regarded as a primordial form 
of the present holographic theory of area-entropy relation$^{[5]}$ or the holographic principle, 
suggested so far in the framework of the present quantum theory subject to the Heisenberg's commutation 
relations that appear now in the Inonu-Wigner's contraction limit$^{[6]}$ of YSTA, $\lambda \rightarrow 0$ and 
$R \rightarrow \infty,$ as was shown in [1].   

At this point, it is worthy of noting the following view,$^{[7]}$ that the holographic 
principle should play the role of guide from the present physical world subject to 
the ordinary quantum theory to that of the ultimate theory of gravitation free from divergences and 
may be fully understood in the latter ultimate theory. It reminds us of the history of black body radiation, 
which originated in classical physics and was fully understood in quantum physics, playing the role of guide from 
the former to the latter physics. We believe that the noncommutative $D_0$ brane field theory described appropriately 
on the Yang's quantized space-time$^{[1],[3]}$ may be a candidate towards such an ultimate theory. 

The present paper is organized as follows. In Sec.\ 2, we briefly recapitulate Yang's quantized space-time 
algebra (YSTA) and its representations. Sec.\ 3 is devoted to review 
the derivation of  $D_0$ brane or $D$-particle field equation subject to YSTA according to the Moyal star 
product formalism for our subsequent consideration. In Sec.\ 4, we find out a certain 
kind of kinematical holographic relation in YSTA on the basis of a definite reduction of spatial degrees of freedom 
in YSTA, compared with the ordinary lattice space. In the final section, Sec.\ 5, the relation between our kinematical 
holographic relation subject to YSTA and the ordinary holographic area-entropy relation is discussed.

\section{\normalsize Yang's Quantized Space-Time Algebra (YSTA) and \break Its Representations}  

\subsection{\normalsize Yang's Quantized Space-Time Algebra (YSTA) }

Let us first recapitulate briefly the Lorentz-covariant Yang's quantized space-time algebra (YSTA).
$D$-dimensional Yang's quantized space-time algebra is introduced$^{[1],[2]}$ as the 
result of the so-called Inonu-Wigner's contraction procedure with two contraction parameters, $R$ and 
$\lambda$, from $SO(D+1,1)$ algebra with generators $\hat{\Sigma}_{MN}$; 
\begin{eqnarray}
 \hat{\Sigma}_{MN}  \equiv i (q_M \partial /{\partial{q_N}}-q_N\partial/{\partial{q_M}}),
\end{eqnarray}
which work on $(D+2)$-dimensional parameter space  $q_M$ ($M= \mu,a,b)$ satisfying  
\begin{eqnarray}
             - q_0^2 + q_1^2 + ... + q_{D-1}^2 + q_a^2 + q_b^2 = R^2.
\end{eqnarray}
 
Here, $q_0 =-i q_D$ and $M = a, b$ denote two extra dimensions with space-like metric signature.

$D$-dimensional space-time and momentum operators, $\hat{X}_\mu$ and $\hat{P}_\mu$, 
with $\mu =1,2,...,D,$ are defined in parallel by
\begin{eqnarray}
&&\hat{X}_\mu \equiv \lambda\ \hat{\Sigma}_{\mu a}
\\
&&\hat{P}_\mu \equiv \hbar /R \ \hat{\Sigma}_{\mu b},   
\end{eqnarray}
together with $D$-dimensional angular momentum operator $\hat{M}_{\mu \nu}$
\begin{eqnarray}
   \hat{M}_{\mu \nu} \equiv \hbar \hat{\Sigma}_{\mu \nu}
\end{eqnarray} 
and the so-called reciprocity operator
\begin{eqnarray}
    \hat{N}\equiv \lambda /R\ \hat{\Sigma}_{ab}.
\end{eqnarray}
Operators  $( \hat{X}_\mu, \hat{P}_\mu, \hat{M}_{\mu \nu}, \hat{N} )$ defined above 
satisfy the so-called contracted algebra of the original $SO(D+1,1)$, or Yang's 
space-time algebra (YSTA):
\begin{eqnarray}
&&[ \hat{X}_\mu, \hat{X}_\nu ] = - i \lambda^2/\hbar \hat{M}_{\mu \nu}
\\
&&[\hat{P}_\mu,\hat{P}_\nu ] = - i\hbar / R^2\ \hat{M}_{\mu \nu}
\\
&&[\hat{X}_\mu, \hat{P}_\nu ] = - i \hbar \hat{N} \delta_{\mu \nu}
\\
&&[ \hat{N}, \hat{X}_\mu ] = - i \lambda^2 /\hbar  \hat{P}_\mu
\\
&&[ \hat{N}, \hat{P}_\mu ] =  i \hbar/ R^2\ \hat{X}_\mu,
\end{eqnarray}
with familiar relations among ${\hat M}_{\mu \nu}$'s omitted.

\subsection{Quasi-Regular Representation of YSTA}

Let us further recapitulate briefly the representation$^{[1],[3]}$ of YSTA for the subsequent 
consideration in section 4. First, it is important to notice the following elementary fact that ${\hat\Sigma}_{MN}$ 
defined in Eq.(2.1) with $M, N$ being the same metric signature have discrete eigenvalues, i.e., $0,\pm 1 ,
\pm 2,\dots$, and those with $M, N$ being opposite metric signature have continuous eigenvalues,
$\footnote{The corresponding eigenfunctions are explicitly given in ref. [3].}$ consistently with 
covariant commutation relations of YSTA. This fact was first emphasized by Yang$^{[2]}$ in connection with 
the preceding Snyder's quantized space-time.$^{[2]}$ This conspicuous aspect is well understood by means of 
the familiar example of the three-dimensional angular momentum in quantum mechanics, where individual components, 
which are noncommutative among themselves, are able to have discrete eigenvalues, consistently with the 
three-dimensional rotation-invariance. 
 
This fact implies that Yang's space-time algebra (YSTA) presupposes for its representation space 
to take representation bases like 
\begin{eqnarray}
|{\hat{\Sigma}}_{0a}= t/\lambda,n...> \equiv |{\hat{\Sigma}}_{0a} =t/\lambda> |{\hat{\Sigma}}_{12}=n_{12}>
...|{\hat{\Sigma}}_{910}=n_{910}>,
\end{eqnarray}
where $t$ clearly denotes {\it time}, the continuous eigenvalue of $\hat{X}_0 \equiv \lambda\ \hat{\Sigma}_{0 a}$ 
and $n...$ discrete eigenvalues of maximal commuting set of subalgebra of $SO(D+1,1)$ which are 
commutative with ${\hat{\Sigma}}_{0a}$, for instance, ${\hat{\Sigma}}_{12}$, ${\hat{\Sigma}}_{34},$..., 
${\hat{\Sigma}}_{910}$, when $D=11$.

Indeed, an infinite dimensional linear space expanded by $|{\hat{\Sigma}}_{0a}= t/\lambda, 
n...>$ mentioned above provides a representation space of unitary infinite dimensional representation of YSTA. It is 
the so-called "quasi-regular representation"$^{[8]}$ of SO(D+1,1),\footnote{It corresponds, in the case of unitary 
representation of Lorentz group $SO(3,1)$, to taking $K_3\ (\sim \Sigma_{03})$ and $J_3\ (\sim \Sigma_{12})$ to be 
diagonal, which have continuous and discrete eigenvalues, respectively, instead of ${\bf J}^2$ and $J_3$ in 
the familiar representation.}
and is decomposed into the infinite series of the ordinary unitary irreducible representations" of 
$SO(D+1,1)$ constructed on its maximal compact subalgebra, $SO(D+1)$. 

It means that there holds the following form of decomposition theorem,
\begin{eqnarray}
|{\hat{\Sigma}}_{0a}= t/\lambda, n...>= \sum_{\sigma 's}\ \sum_{l,m}\  C^{\sigma's, n... }_{l,m}(t/\lambda)\ 
 | \sigma 's ; l,m>,
\end{eqnarray}      
with expansion coefficients $C^{\sigma's, n... }_{l,m}(t/\lambda).$ In Eq.(2.13), 
$|\sigma 's ; l, m>'s$ on the right hand side describe the familiar unitary irreducible representation 
bases of $SO(D+1,1)$, which are designated by $\sigma 's$ and $(l,m),$  
\footnote{In the familiar unitary irreducible representation of $SO(3,1)$, it is well known that $\sigma$'s are 
represented by two parameters, $(j_0, \kappa)$, with $j_0$ being $1,2,...\infty$ and $\kappa$ being purely 
imaginary number, for the so-called principal series of representation. With respect to the associated representation 
of $SO(3)$, when it is realized on $S^2$, as in the present case, $l$'s denote positive integers, 
$l= j_0, j_0+1, j_0+2,...,\infty$, and $m$ ranges over $\pm l, \pm(l-1) ,...,\pm1, 0.$ } 
denoting, respectively, the irreducible unitary representations of $SO(D+1,1)$ and the associated 
irreducible representation bases of $SO(D+1)$, the maximal compact subalgebra of $SO(D+1,1)$, mentioned above. 
It should be noted here that, as will be remarked in section 4, $l$'s  are limited to be integer, 
excluding the possibility of half-integer, because of the fact that generators of $SO(D+1)$ in YSTA are 
defined as differential operators on $S^D$, 
i.e., ${q_1}^2 + {q_2}^2 + ... + {q_{D-1}}^2 + {q_a}^2 + {q_b}^2 = 1.$

In what follows, let us call the infinite dimensional representation space introduced above for the representation of 
YSTA, Hilbert space I, in distinction to Hilbert space II which is Fock-space constructed dynamically by 
creation-annihilaltion operators of second-quantized fields on YSTA, such as D-particle field,$^{[11]}$ discussed 
in section 5.
 
\section{\normalsize $D_0$ Brane Field Equation in Moyal Star Product Formalism of YSTA }

In order to search for a possible holographic principle in YSTA, let us first remember the derivation of 
the field equation of ${D_0}$ brane on the Yang's quantized space-time given in ref. [3], by making 
use of the method of Moyal star product. We assumed there the $D_0$ brane field 
to be the function of ${\Sigma}_{K a}$ with $K$ ranging over $\mu$ and $b$, that is, 
$D({\Sigma}_{K a})= D(X_\mu, N)$ which is the minimal function to admit the infinitesimal translation 
operation, i.e., $X_\mu \rightarrow X_\mu + \alpha_\mu N,\ N \rightarrow N -\alpha_\mu X_\mu /R^2.$

Furthermore, ${\Sigma}_{MN}$ 
corresponding to $ \hat{\Sigma}_{MN}$ defined in Eq.(2.1) is now given as the function of 
the classical canonical variables, $q_M$ and $p_M$, that is ,
\begin{eqnarray} 
{\Sigma}_{MN} = ( - q_M p_N  + q_N p_M).
\end{eqnarray}

As was stated in ref.[3], the Moyal star product of any two functions of $\Sigma_{MN}(q,p)$,
$F(\Sigma) * G(\Sigma),$ is by definition given in the following covariant form,
\begin{eqnarray}
F (\Sigma) * G(\Sigma)  = F (\Sigma )\ exp\ {i \over 2}\ (  \overleftarrow{\partial/\partial{\Sigma_{MN}}}\ 
{\Sigma }_{NO}\ \overrightarrow{\partial/{\partial {\Sigma}_{OM}}})\ G (\Sigma) .
 \end{eqnarray}

Starting from the following hypothetical action of $D_0$ brane field, $\bar {\hat {L}}$, after the pioneer work 
``M-theory as a matrix model,"$^{[9]}$ 

\begin{eqnarray}
&&\bar{\hat L}  = A\ {\rm tr}\  \{ [\hat {\Sigma}_{KL}, \hat {D}^\dagger]\ [\hat {\Sigma}_{KL}, \hat {D}]\ \}  
\nonumber\\
   && \hskip0.5cm = A'\ {\rm tr}\  \{ 2\ (R^2 /\hbar^2)\  [{\hat P}_\mu, \hat {D}^\dagger ]\ [ \hat {P}_\mu, \hat {D}]
\nonumber\\ 
 && \hskip2cm - {\lambda}^{-4 }\ [\ [\hat {X}_\mu, \hat {X}_\nu],\ \hat {D}^\dagger]\ [ [\hat {X}_\mu,\hat {X}_\nu], 
\hat {D}]\ \}.
\end{eqnarray}
with $K,L$ ranging over $(\mu, b),$ 
and replacing the commutator $[\hat {F}, \hat {G}]$ by the Moyal bracket $[F,G]_M \equiv 
F\star G - G \star F$ and ${\rm tr}$ by the integrations over $X_\mu, N$ in the above expression,
we arrived at the following $D_0$ brane field equation$^{[3]}$
\begin{eqnarray} 
&&[\ {\Sigma_{aK}}^2\ (\partial / {\partial \Sigma_{aL}})^2  -(\Sigma_{aK}  \partial/\partial {\Sigma_{aK}})^2 
\nonumber\\
&& \hskip2.5cm - (D-1) \Sigma_{aK}  \partial /{\partial \Sigma_{aK}}\ ]\ D (\Sigma_{aM})= 0,
\nonumber
\end{eqnarray}
or
\begin{eqnarray}
&&[( {X_\sigma}^2 +R^2 N^2 ) ( (\partial/\partial {X_\mu})^2 + R^{-2} (\partial/\partial {N})^2)) 
 - ( X_\mu \partial/\partial {X_\mu} + N \partial/\partial{N})^2 
\nonumber\\
&&\hskip1cm - (D-1) ( X_\mu \partial/\partial {X_\mu}+N \partial/\partial{N})\ ]\ D ( X_\nu, N) = 0.
\end{eqnarray} 

One should notice that the above equation is certainly a $(D+1)$-dimensional field equation with 
coordinates $X_\mu, R N$ and tends to the $D$-dimensional massless field equation in the 
Inonu-Wigner's contraction limit of YSTA, that is, $\lambda \rightarrow 0$ and $R \rightarrow \infty,$ 
as was noticed in ref.[1].

\section{\normalsize Kinematical Holographic Relation in YSTA}

Now, under the preliminaries given in the preceding sections, let us try to find a possible origin of area-entropy relation 
in YSTA, taking in mind the structure of the above field equation of $D_0$ brane, Eq.(3.4), with $(D+1)$-dimensional 
coordinates, $X_\mu$ and $R N$.  

A naive consideration leads us to examining the {\it kinematical} holographic relation between the area of boundary 
surface of any $D$-dimensional space-like region with radius $L$ in the above $(D+1)$-dimensional coordinate space 
and the number of spatial degrees of freedom inside this space-like region. 

Let us define the above boundary surface of the $D$-dimensional space-like 
region with finite radius {L} in the unit of $\lambda$ by
\begin{eqnarray}
 \sum_{K \neq 0}{\Sigma_{aK}}^2  = \sum_{\mu \neq 0}{\Sigma_{a \mu}}^2 + {\Sigma_{ab}}^2 = (L/\lambda)^2,
\end{eqnarray}
or
\begin{eqnarray}
{X_1}^2 + {X_2}^2 + \cdot + {X_{D-1}}^2 + R^2\ N^2 = L^2.
\end{eqnarray}

The area of the above region, ${\cal A}$ (in the unit of length ${\lambda}),$ is therefore nothing but the area of 
$S^{D-1}$, that is, the $(D-1)$-dimensional spherical surface with radius $L/\lambda$, which is given 
by
\begin{eqnarray}
{\cal A} =  ({\rm area\ of}\ S^{D-1}) ={(2 \pi)^{D/2} \over {(D-2)!!}} (L/\lambda)^{D-1}, 
&&for\ D\ even
\nonumber\\                   =2 {(2\pi)^{(D-1)/2} \over {(D-2)!!}} (L/\lambda)^{D-1}. &&for\ D\ odd 
\end{eqnarray}
 
Next, let us count the number of spatial degrees of freedom in this bounded region with radius $L$, 
denoted by $n^L_{\rm dof}.$  In order to calculate $n^L_{\rm dof}$ in YSTA, however, one should notice the following 
important fact that spatial quantities in YSTA are noncommutative operators, $ \hat{X}_u$ and $(R) \hat{N}$, 
subject to YSTA and consequently $n^L_{\rm dof}$ should be, logically and also practically, found in the {\it 
representation space} of YSTA, called Hilbert space I defined in subsection 2.2.  

Indeed, one finds that the representation space needed to calculate $n^L_{\rm dof}$ is 
prepared in Eq.(2.13), where any "quasi-regular" representation basis,\\ $ |{\hat{\Sigma}}_{0a}= t/\lambda, n...>$, is 
decomposed into the infinite series of the ordinary unitary representation bases of $SO(D+1,1)$, $| \sigma 's ; l,m>.$ 
As was stated in subsection 2.2, the latter representation bases, $| \sigma 's ; l,m>'s$ are constructed on 
the familiar finite dimensional representations of maximal compact subalgebra of YSTA, $SO(D+1)$, whose representation 
bases are labeled by $(l,m)$ and provide the representation bases for spatial quantities under consideration, 
because $SO(D+1)$ just involves those spatial operators $( \hat{X}_u, R \hat{N})$. 

In order to arrive at the final goal of counting $n^L_{\rm dof}$, therefore, one has only to find mathematically 
a certain irreducible representation of $SO(D+1)$, which {\it properly} describes (as seen in what follows) 
the spatial quantities $( \hat{X}_u, R \hat{N})$ inside the bounded region with radius $L$, then one finds 
$n^L_{\rm dof}$ through counting the dimension of the representation. 

At this point, it is important to note that, as was remarked in advance in subsection 2.2, any generators of $SO(D+1)$ 
in YSTA  are defined by the differential operators on the $D-$dimensional unit sphere, $S^D$, i.e., 
${q_1}^2 + {q_2}^2 + ... + {q_{D-1}}^2 + {q_a}^2 + {q_b}^2 = 1,$ limiting its representations with $l$ to be 
integer.
   
On the other hand, it is well known that the irreducible representation of arbitrary high-dimensional $SO(D+1)$ 
on $S^D = SO(D+1)/SO(D)$ is derived in the algebraic way, $^{[10]}$ irrelevantly to any detailed knowledge of 
the decomposition equation (2.13), but solely in accord with the fact that $SO(D+1)$ in YSTA is defined originally 
on $S^D$, as mentioned above. One can choose, for instance, $SO(D)$ with generators $\hat{\Sigma}_{MN} (M,N=b, u)$, 
while $SO(D+1)$ with generators $\hat{\Sigma}_{MN}(M,N=a,b,u)$. Then, it turns out that any irreducible representation 
of $SO(D+1)$, denoted by $\rho_l$, is uniquely designated by the maximal integer $l$ of eigenvalues of 
${\hat \Sigma}_{ab}$ in the representation, where ${\hat \Sigma}_{ab}$ is known to be a possible Cartan subalgebra 
of the so-called compact symmetric pair $(SO(D+1),SO(D))$ of rank $1$.$^{[10]}$ 

According to the so-called Weyl's dimension formula, the dimension of $\rho_l$ is given by$^{[10],[1]}$
\begin{eqnarray}
 dim\ (\rho_l)= {(l+\nu) \over \nu} {(l+2\nu-1)! \over {l!(2\nu -1)!}},
\end{eqnarray}
where $ \nu \equiv (D-1)/2$ and $D \geq 2$.\footnote{This equation just gives the familiar result $dim\ (\rho_l)= 2l +1,$ 
in the case $SO(3)$ taking $D=2.$} 

At this point, let us find a certain irreducible representation of $SO(D+1)$ among those $\rho_l 's $  given above, 
which {\it properly} describes (or realizes) the spatial quantities inside the bounded region with radius $L$. 
Now, let us choose tentatively $l = [L/\lambda]$ with $[L/\lambda]$ being the integer part of $L/\lambda$. In this case, 
one finds out that 
the representation $\rho_{[L/\lambda]}$ just {\it properly} describes all of generators of $SO(D+1)$ inside the above 
bounded spatial region with radius $L$, because  $[L/\lambda ]$ indicates also the largest eigenvalue of any generators 
of $SO(D+1)$ in the representation $\rho_{[L/\lambda]}$ on account of its $SO(D+1)-$invariance and 
hence eigenvalues of spatial quantities $( \hat{X}_u, R \hat{N})$ are well confined inside the bounded region with 
radius {L}. 
As the result, one finds that the dimension of $\rho_{[L/\lambda]}$ just gives the number of spatial degrees of 
freedom inside this bounded region, $n^L_{\rm dof}$, according to the preceding argument.

In consequence, one finds 
\begin{eqnarray}
&&n^L_{\rm dof}  =  dim\ ( \rho_{[L/\lambda]}) = {2 \over (D-1)!}{([L/\lambda]+ D-2)! \over ([L/\lambda]-1)!}
\nonumber\\
 &&\hskip3.5cm \sim {2 \over (D-1)!}  [L/\lambda]^{D-1},
\end{eqnarray}taking into account the fact that $[L/\lambda] \gg D.$ 

At this point, by comparing both Eqs.(4.3) and (4.5), we find out a very important relation, 
that is, the proportional relation between ${\cal A}$ and $n^L_{\rm dof},$    
\begin{eqnarray} 
n^L_{\rm dof}= {\cal A} / G,
\end{eqnarray}
with the proportional constant $G$ given by
\begin{eqnarray}
G\  ( = {\cal A}/n^L_{\rm dof})\ \sim {(2 \pi)^{D/2} \over 2}\ (D-1)!! &&for\ D\ even
\\
    \sim  (2 \pi)^{(D-1)/2}(D-1)!! &&for\ D\ odd,
\end{eqnarray}
which essentially depends only on the dimension $D$. 

Let us call this relation given by Eq.(4.6) the kinematical holographic relation in Yang's 
space-time algebra (YSTA) on account of its definite kinematical nature, as seen in the above 
derivation. It may be regarded as a primordial form of the so-called holographic principle 
widely studied in the framework of the present quantum theory. 

Before investigating the relation between the kinematical holographic relation, Eq.(4.6), and the 
ordinary holographic 
relation, let us consider the corresponding relation in the case of the familiar $D$-dimensional 
commutative lattice space with lattice spacing $\lambda$. In this case, the area ${\cal A}$ of 
$(D-1)$-dimensional boundary surface is given by Eq.(4.3) as it stands, while $[n]^L_{\rm dof}$, the 
number of spatial degrees of 
freedom in the commutative lattice space mentioned above is given by the well-known $D$-dimensional 
volume with radius $L/\lambda$, that is,
\begin{eqnarray}
[n]^L_{\rm dof} = {(2\pi)^{D/2} \over D!!} (L/\lambda)^D  &&for\ D\ even,
\nonumber\\
            = {2(2\pi)^{(D-1)/2} \over D!!} (L/\lambda)^D &&for\ D\ odd.
\end{eqnarray}

Comparing this result with Eq.(4.5), we find out a marked 
difference between $n^L_{\rm dof}$ in YSTA and $[n]^L_{\rm dof}$ in the lattice space, that is, a definite 
reduction of spatial degrees of freedom in YSTA in comparison with the case of lattice space, which 
is clearly due to the noncommutativity of YSTA and enables us to arrive at the kinematical 
holographic relation, Eq.(4.6).

\section{\normalsize Concluding Remarks }

In the preceding section, we have found the kinematical holographic relation in YSTA, Eq.(4.6).   
The relation should be further compared with the ordinary holographic relation$^{[5]}$
\begin{eqnarray}
N_{\rm dof} \leq  {\cal A }/ 4.
\end{eqnarray}
Here $N_{\rm dof}$ by definition means the total number of independent quantum degrees of freedom or the number 
of orthonormal bases of Hilbert space II needed to describe all physics inside the 
bounded spatial region with the boundary area ${\cal A}$ (in the unit of Planck length ${l_P}$), although 
$n^L_{\rm dof}$ in Eq.(4.6) refers to Hilbert space I, according to the definition of Hilbert space 
I, II given in subsection 2.2.

The direct derivation of the holographic relation Eq.(5.1) from the kinematical holographic 
relation Eq.(4.6), however, turns out to be principally difficult, because the former relation is viewed 
to hold in the physical 
world subject to the ordinary quantum theory, while the latter relation to hold in the world 
subject to YSTA. As a matter of fact, the former world is certainly kinematically derivable from the latter 
through the Inonu-Wigner contraction limit, $\lambda \rightarrow 0$ and $R \rightarrow \infty,$ and the 
value of coefficient factor $G$ in Eq. (4.6) remains invariant under this limit, but $n^L_{\rm dof}$ 
and ${\cal A}$ on both sides, clearly tend to infinity.

Therefore, we have to step forward to calculate dynamically $N^L_{\rm dof}$, i.e., the number of 
independent field degrees of freedom subject to YSTA, such as $D_0$ brane field inside the bounded 
spatial region with the number of spatial degrees of freedom, $n^L_{\rm dof}.$ 

At this point, if we take into consideration that $D_0$ brane field in YSTA is certainly a kind of 
$n^L_{\rm dof} \times n^L_{\rm dof}$ matrix field with respect to Hilbert space I, we are likely to 
imagine that there occurs a large number of contribution to $N^L_{\rm dof}$, of the order of 
$(n^L_{\rm dof})^2$. This consideration, however, turns out to be premature if one takes 
into account the fact that $D$-particles as the fundamental constituents or partons of all systems can 
never be observables with individuality as the objects of holographic view.

Consequently, in order to answer this problem finally, we have to deal more in detail with 
the physical system such as black hole$^{[12]}$ on the basis of supersymmetric extension of $D_0$ brane 
action, given by Eq.(3.3). The problem, however, must be left to the forthcoming paper.

\end{document}